\documentclass[12pt,3p]{elsarticle}
\usepackage{graphicx}
\usepackage{amssymb}
\usepackage{url}

\usepackage{listings}
\usepackage{xcolor}
\usepackage{amssymb}
\usepackage{epstopdf}
\usepackage{color}
\usepackage{fancyhdr}
\usepackage{lastpage}
\usepackage{longtable}
\usepackage{array}
\usepackage{subfigure}
\usepackage{breakurl}
\graphicspath{{./Figs/}}
\usepackage{amsmath}
\usepackage{algorithm}
\usepackage{algorithmic}
\usepackage{multirow}
\usepackage[colorlinks,linkcolor=blue,anchorcolor=blue,citecolor=blue]{hyperref}

\newcommand{\tabincell}[2]{\begin{tabular}{@{}#1@{}}#2\end{tabular}}

\begin{document}

\begin{frontmatter}
\title{A Survey on the Security of Blockchain Systems}

\author[PolyU]{Xiaoqi~Li}
\ead{csxqli@gmail.com}
\author[PolyU]{Peng~Jiang}
\author[uest]{Ting~Chen}
\author[PolyU]{Xiapu~Luo\corref{cor1}}
\ead{csxluo@comp.polyu.edu.hk}\cortext[cor1]{Corresponding author}
\author[bupt]{Qiaoyan~Wen}

\address[PolyU]{Department of Computing, The Hong Kong Polytechnic University, Hong Kong SAR}
\address[uest]{Center for Cybersecurity, University of Electronic Science and Technology of China, China}
\address[bupt]{State Key Laboratory of Networking and Switching Technology, Beijing University of Posts and Telecommunications, China}

\begin{abstract}
\label{sec:abstract}
Since its inception, the blockchain technology has shown promising application prospects. From the initial cryptocurrency to the current smart contract, blockchain has been applied to many fields. Although there are some studies on the security and privacy issues of blockchain, there lacks a systematic examination on the security of blockchain systems. In this paper, we conduct a systematic study on the security threats to blockchain and survey the corresponding real attacks by examining popular blockchain systems. We also review the security enhancement solutions for blockchain, which could be used in the development of various blockchain systems, and suggest some future directions to stir research efforts into this area.
\end{abstract}


\end{frontmatter}

\section{Introduction}
\label{sec:introduction}

Since the debut of Bitcoin in 2009, its underlying technique, blockchain, has shown promising application prospects and attracted lots of attentions from academia and industry. Being the first cryptocurrency, Bitcoin was rated as the top performing currency in 2015~\cite{2} and the best performing commodity in 2016~\cite{3}, and has more than 300K confirmed transactions~\cite{4} daily in May, 2017. At the same time, the blockchain technique has been applied to many fields, including medicine~\cite{ekblaw2016case,azaria2016medrec,yue2016healthcare}, economics~\cite{huckle2016internet,bylica2015probabilistic,hurich2016virtual}, Internet of things~\cite{dorri2017blockchain,zhang2016iot,sun2016blockchain}, software engineering~\cite{xu2016blockchain,nordstrom2015personal,czepluch2015use} and so on. The introduction of Turing-complete programming languages to enable users to develop smart contracts running on the blockchain marks the start of blockchain 2.0 era. With the decentralized consensus mechanism of blockchain, smart contracts allow mutually distrusted users to complete data exchange or transaction without the need of any third-party trusted authority. Ethereum is now (May of 2017) the most widely used blockchain supporting smart contracts, where there are already 317,506 smart contracts and more than 75,000 transactions happened daily~\cite{6}.

Since blockchain is one of the core technology in FinTech (Financial Technology) industry, users are very concerned about its security. Some security vulnerabilities and attacks have been recently reported. Loi et al. discover that 8,833 out of 19,366 existing Ethereum contracts are vulnerable~\cite{luu2016making}. Note that smart contracts with security vulnerabilities may lead to financial losses. For instance, in June 2016, the criminals attacked the smart contract \texttt{DAO} ~\cite{7} by exploiting a recursive calling vulnerability, and stole around 60 million dollars.
As another example, in March 2014, the criminals exploited transaction mutability in Bitcoin to attack \texttt{MtGox}, the largest Bitcoin trading platform. It caused the collapse of \texttt{MtGox}, with a value of 450 million dollars Bitcoin stolen~\cite{8}.

Although there are some recent studies on the security of blockchain, none of them performs a systematic examination on the risks to blockchain systems, the corresponding real attacks, and the security enhancements. The closest research work to ours is~\cite{atzeisurvey} that only focuses on Ethereum smart contracts, rather than popular blockchain systems. From security programming perspective, their work analyzes the security vulnerabilities of Ethereum smart contracts, and provides a taxonomy of common programming pitfalls that may lead to vulnerabilities~\cite{atzeisurvey}. Although a series of related attacks on smart contracts are listed in~\cite{atzeisurvey}, there lacks a discussion on security enhancement. This paper focuses on the security of blockchain from more comprehensive perspectives. The main contributions of this paper are as follows:

(1). To the best of our knowledge, we conduct the \textit{first} systematic examination on security risks to popular blockchain systems.

(2). We survey the real attacks on popular blockchain systems from 2009 to the present (May of 2017) and analyze the vulnerabilities exploited in these cases.

(3). We summarize practical academic achievements for enhancing the security of blockchain, and suggest a few future directions in this area.

The remainder of this paper is organized as follows. Section~\ref{sec:background} introduces the main technologies used in blockchain systems. Section~\ref{sec:risks} systematically examines the security risks to blockchain, and Section~\ref{sec:attackcases} surveys real attacks on blockchain systems. After summarizing the security enhancements to blockchain in Section~\ref{sec:enhancement}, we suggest a few future directions in Section~\ref{sec:future}. Finally, Section~\ref{sec:conclusion} concludes the paper.

\section{Overview of Blockchain Technologies}
\label{sec:background}
This section introduces the main technologies employed in blockchain. We first present the fundamental trust mechanism (i.e., the consensus mechanism) used in blockchain, and then explain the synchronization process between nodes. After that, we introduce the two development stages of blockchain.

\subsection{Consensus Mechanism}
Being a decentralized system, blockchain systems do not need a third-party trusted authority. Instead, to guarantee the reliability and consistency of the data and transactions, blockchain adopts the decentralized consensus mechanism. In the existing blockchain systems, there are four major consensus mechanisms~\cite{zheng2016blockchain}: PoW (Proof of Work), PoS (Proof of Stake), PBFT (Practical Byzantine Fault Tolerance), and DPoS (Delegated Proof of Stake). Other consensus mechanisms, such as PoB (Proof of Bandwidth)~\cite{11}, PoET (Proof of Elapsed Time)~\cite{12}, PoA(Proof of Authority)~\cite{poa} and so on, are also used in some blockchain systems. The two most popular blockchain systems (i.e., Bitcoin and Ethereum) use the PoW mechanism. Ethereum also incorporates the PoA mechanism (i.e., Kovan public test chain~\cite{Kovan}), and some other cryptocurrencies also use the PoS mechanism, such as PeerCoin, ShadowCash and so on.

\begin{figure}[ht]
	\centering
	\includegraphics[width=3in]{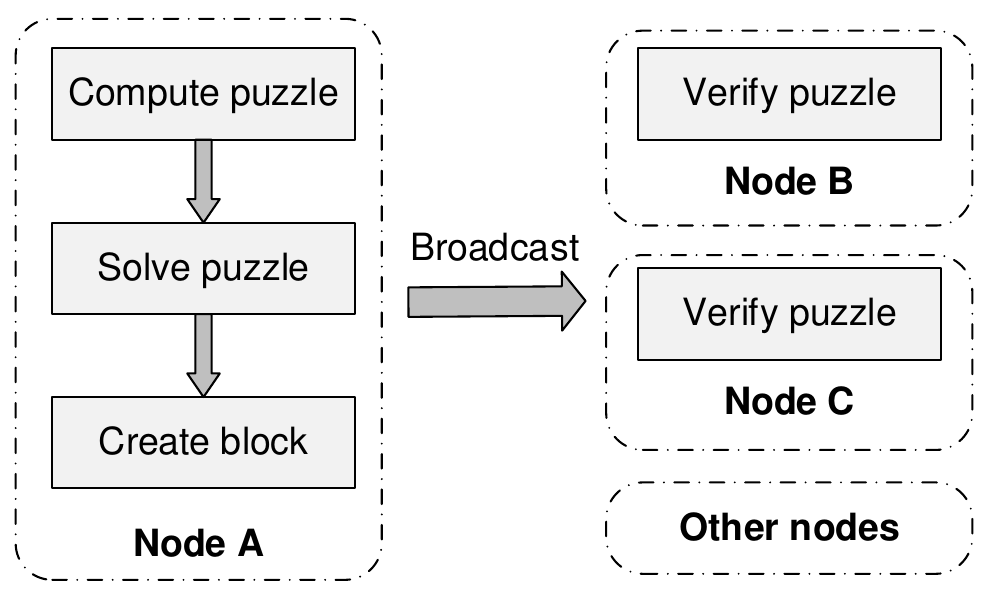}
	\caption{PoW consensus mechanism}
	\label{PoW}
\end{figure}

PoW mechanism uses the solution of puzzles to prove the credibility of the data. The puzzle is usually a computationally hard but easily verifiable problem. When a node creates a block, it must resolve a PoW puzzle. After the PoW puzzle is resolved, it will be broadcasted to other nodes, so as to achieve the purpose of consensus, as shown in Fig.\ref{PoW}.

In different blockchain systems, the block structure may vary in detail. Typically in Bitcoin, each block contains \texttt{PrevHash}, \texttt{nonce}, and \texttt{Tx}~\cite{luu2017smart}. In particular, \texttt{PrevHash} indicates the hash value of the last generated block, and  \texttt{Tx}s denote the transactions included in this block. The value of \texttt{nonce} is obtained by solving the PoW puzzle. A correct \texttt{nonce} should satisfy that the hash value shown in Equation~\ref{PoWequation} is less than a target value, which could be adjusted to tune the difficulty of PoW puzzle.

\begin{equation}
SHA256(PrevHash\left |  \right |Tx1\left |  \right |Tx2\left |  \right | . . . \left |  \right |nonce) < Target
\label{PoWequation}
\end{equation}

PoS mechanism uses the proof of ownership of cryptocurrency to prove the credibility of the data. In PoS-based blockchain, during the process of creating block or transaction, users are required to pay a certain amount of cryptocurrency. If the block or transaction created can eventually be validated, the cryptocurrency will be returned to the original node as a bonus. Otherwise, it will be fined. In the PoW mechanism, it needs a lot of calculation, resulting in a waste of computing power. On the contrary, PoS mechanism can greatly reduce the amount of computation, thereby increasing the throughput of the entire blockchain system.

\subsection{Block Propagation and Synchronization}

\begin{figure}[ht]
	\centering
	\includegraphics[width=4in]{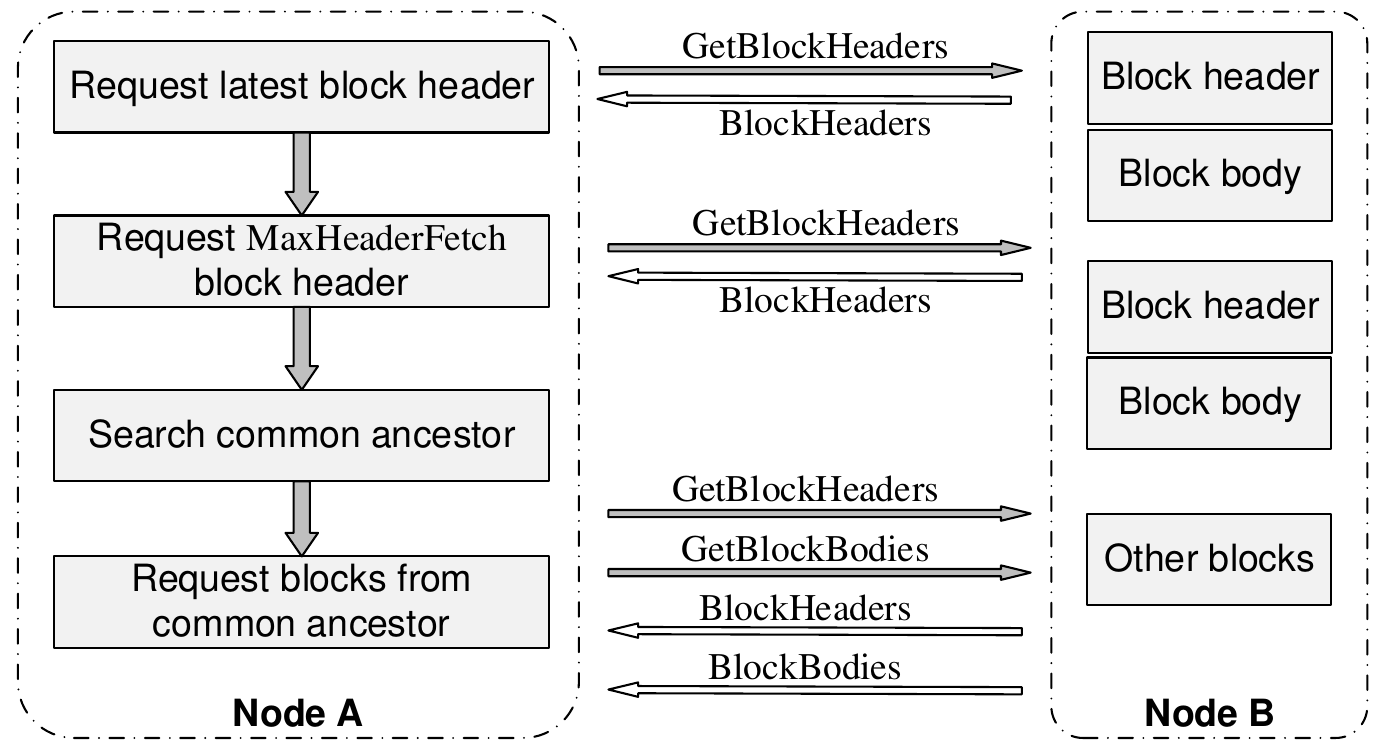}
	\caption{Block synchronization process between nodes}
	\label{synchronisation}
\end{figure}

In the blockchain, each full node stores the information of all blocks. Being the foundation to building consensus and trust for blockchain, the block propagation mechanisms can be divided into the following categories~\cite{wa14st2016security,wa14st2016ethereum,gervaissecurity}:

(1). Advertisement-based propagation. This propagation mechanism is originated from Bitcoin. When node \texttt{A} receives the information of a block, \texttt{A} will send an \texttt{inv} message (a message type in Bitcoin) to its connected peers. When node \texttt{B} receives the \texttt{inv} message from \texttt{A}, it will do as follows. If node \texttt{B} already has the information of this block, it will do nothing. If node \texttt{B} does not have the information, it will reply to node \texttt{A}. When node \texttt{A} receives the reply message from node \texttt{B}, node \texttt{A} will send the complete information of this block to node \texttt{B}.
	
(2). Sendheaders propagation. This propagation mechanism is an improvement to the advertisement-based propagation mechanism. In the sendheaders propagation mechanism, node \texttt{B} will send a \texttt{sendheaders} message (a message type in Bitcoin) to node \texttt{A}. When node \texttt{A} receives the information of a block, it will send the block header information directly to node \texttt{B}. Compared with the advertisement-based propagation mechanism, node \texttt{A} does not need to send \texttt{inv} messages, and hence it speeds up the block propagation.
	
(3). Unsolicited push propagation. In the unsolicited push mechanism, after one block is mined, the miner will directly broadcast the block to other nodes. In this propagation mechanism, there is no \texttt{inv} message and \texttt{sendheaders} message. Compared with the previous two propagation mechanisms, unsolicited push mechanism can further improve the speed of block propagation.
	
(4). Relay network propagation. This propagation mechanism is an improvement to the unsolicited push mechanism. In this mechanism, all the miners share a transaction pool. Each transaction is replaced by a global ID, which will greatly reduce the broadcasted block size, thereby further reducing the network load and improving the propagation speed.
	
(5). Push/Advertisement hybrid propagation. This hybrid propagation mechanism is used in Ethereum. We assume that node \texttt{A} has n connected peers. In this mechanism, node \texttt{A} will push the block to $\sqrt{n}$ peers directly. For the other $n-\sqrt{n}$ connected peers, node \texttt{A} will advertise the block hash to them.

Different blockchain systems may use diverse block synchronization processes. In Ethereum, node \texttt{A} can request block synchronization from node \texttt{B} with more total difficulty. The specific process is as follows (shown in Fig.\ref{synchronisation})~\cite{wa14st2016security,wa14st2016ethereum,gervaissecurity}:

(1). Node \texttt{A} requests the header of the latest block from node \texttt{B}. This action is implemented by sending a \texttt{GetBlockHeaders} message. Node \texttt{B} will reply to node \texttt{A} a \texttt{BlockHeaders} message that contains the block header requested by \texttt{A}.

(2). Node \texttt{A} requests \texttt{MaxHeaderFetch} blocks to find common ancestor from node \texttt{B}. The default value of \texttt{MaxHeaderFetch} is 256, but the number of block headers sent by node \texttt{B} to \texttt{A} can be less than this value.

(3). If \texttt{A} has not found common ancestor after the above two steps, node \texttt{A} will continue to send \texttt{GetBlockHeaders} message, requesting one block header each time. Moreover, \texttt{A} repeats in binary search to find the common ancestor in its local blockchain.

(4). After node \texttt{A} discovers a common ancestor, \texttt{A} will request block synchronization from the common ancestor. In this process, \texttt{A} requests \texttt{MaxHeaderFetch} blocks per request, but the actual number of nodes sent from \texttt{B} to \texttt{A} can be less than this value.

\begin{figure}[ht]
	\centering
	\vspace*{-1ex}
	\begin{minipage}[t]{0.3\linewidth}
		\centering
		\includegraphics[width=1.4in]{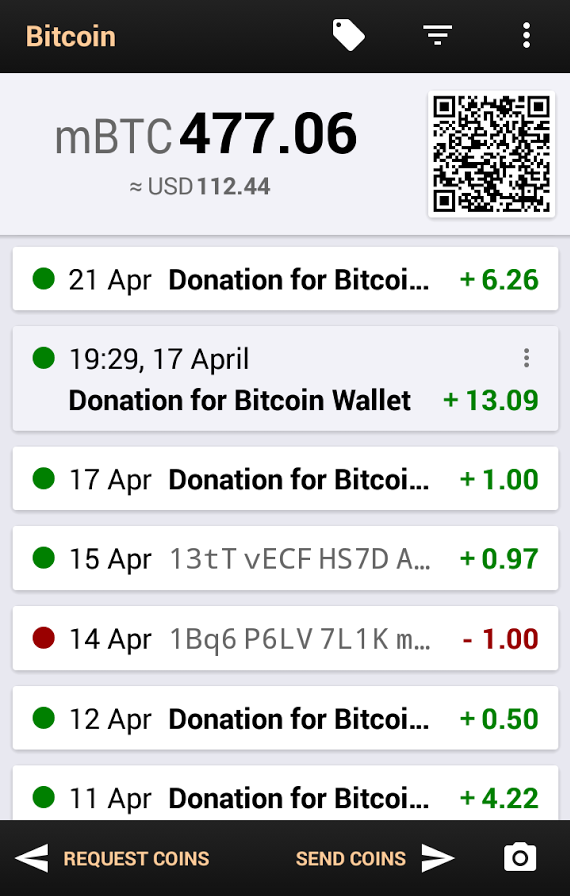}
		\caption{Query Bitcoin transaction history}
		\label{Wallet1}
	\end{minipage}%
	\begin{minipage}[t]{0.3\linewidth}
		\centering
		\includegraphics[width=1.4in]{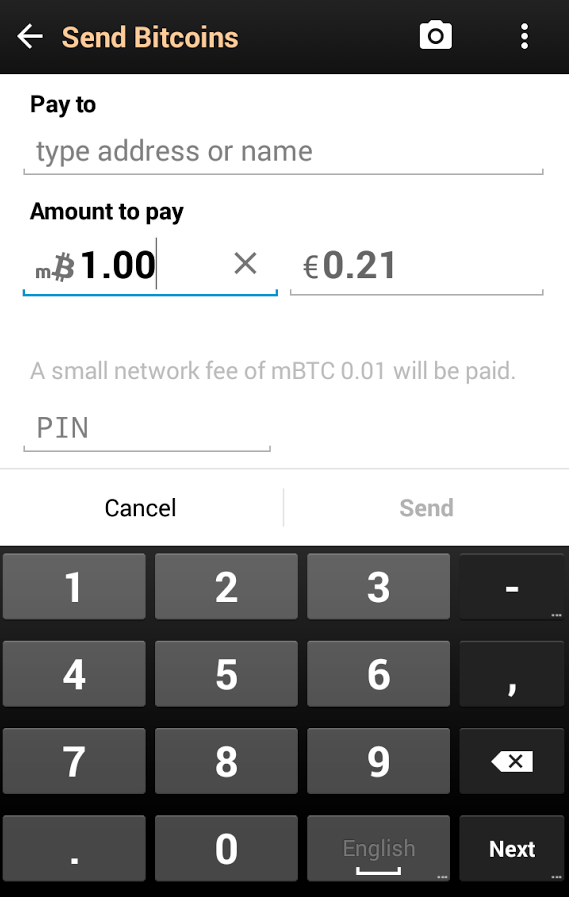}
		\caption{Pay with Bitcoin}
		\label{Wallet2}
	\end{minipage}
	\begin{minipage}[t]{0.3\linewidth}
		\centering
		\includegraphics[width=1.4in]{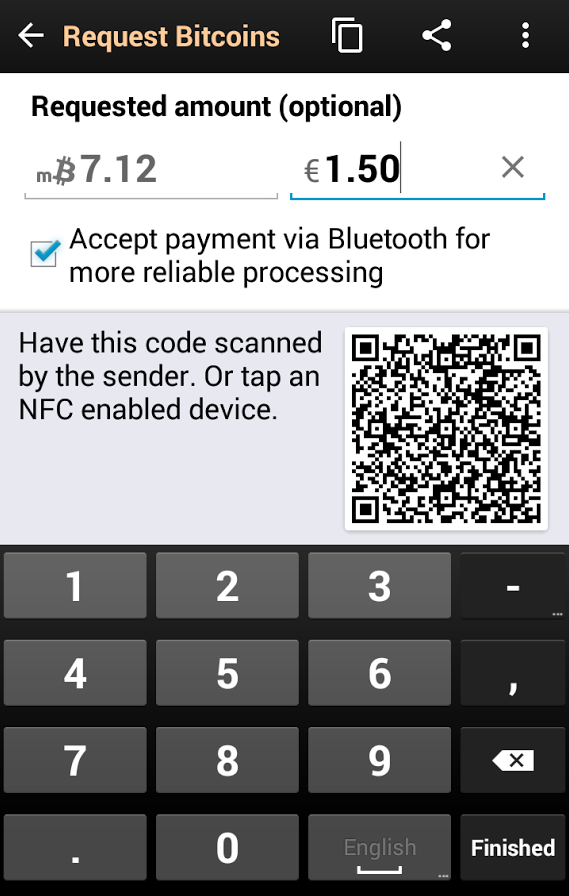}
		\caption{Collect payments with Bitcoin}
		\label{Wallet3}
	\end{minipage}
\end{figure}

\subsection{Technology Development}

From the birth of the first blockchain system Bitcoin, the blockchain technology has experienced two stages of development: blockchain 1.0 and blockchain 2.0.

In the blockchain 1.0 stage, the blockchain technology is mainly used for cryptocurrency. In addition to Bitcoin, there are many other types of cryptocurrencies, such as Litecoin, Dogecoin and so on. There are currently over 700 types of cryptocurrencies, and the total market capitalizations of them are over 26 billion US\$~\cite{19}. The technology stack of cryptocurrency could be divided into two layers: the underlying decentralized ledger layer and protocol layer~\cite{swan2015blockchain}. Cryptocurrency client, such as Bitcoin Wallet~\cite{BitcoinWallet}, runs in the protocol layer to conduct transactions, as shown in Fig.\ref{Wallet1} to Fig.\ref{Wallet3}. Compared with traditional currency, cryptocurrency has the following characteristics and advantages~\cite{21}:

(1). Irreversible and traceable. Transfer and payment operations are irreversible using cryptocurrency. Once the behavior is completed, it is impossible to withdraw. In addition, all user behaviors are traceable, and these behaviors are permanently saved in the blockchain.

(2). Decentralized and anonymous. There is no third-party organization involved in the entire structure of cryptocurrency, nor does it has central management like banks. In addition, all user behaviors are anonymous. Hence, according to the transaction information, we cannot obtain the user's real identity.
	
(3). Secure and permissionless. The security of the cryptocurrency is ensured by the public key cryptography and the blockchain consensus mechanism, which are hard to be broken by the criminal. Moreover, there is no need to apply for any authority or permission to use cryptocurrency. Users can simply use the cryptocurrency through the relevant clients.
	
(4). Fast and global. Transactions can be completed in only several minutes using cryptocurrency. Since cryptocurrencies are mostly based on public chains, anyone in the world can use them. Therefore, the user's geographical location has little impact on the transaction speed.

\begin{figure}[ht]
	\centering
	\includegraphics[width=4.6in]{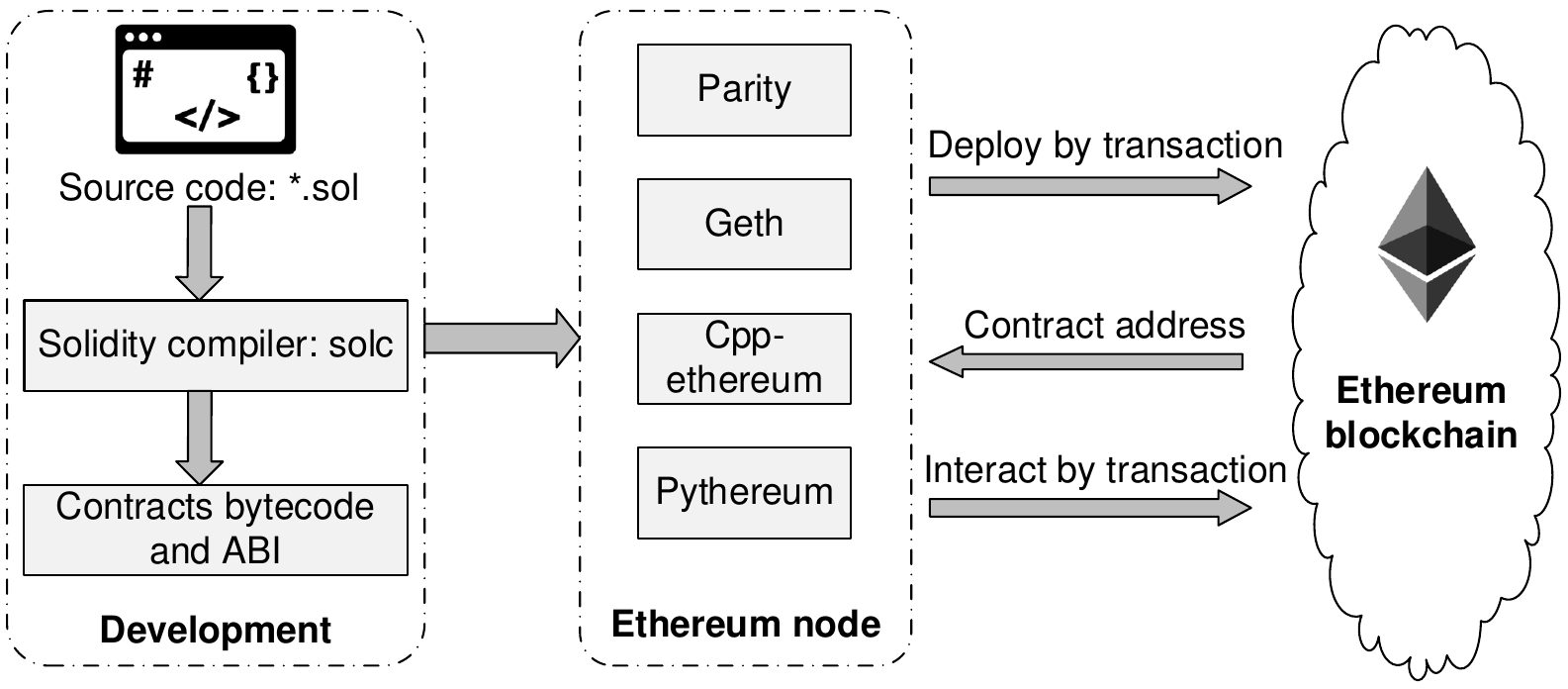}
	\caption{The process of smart contract's development, deployment, and interaction}
	\label{dapp}
\end{figure}

\begin{table}[ht!]
	\centering
	\vspace*{-1ex}
	\scriptsize
	\caption{Statistics of blockchain systems supporting smart contracts (until May of 2017)}
	\vspace{1ex}
	\label{smart}
	\begin{tabular}{|c|c|c|c|}
		\hline
		\textbf{System}&\textbf{Contract language} & \textbf{Total TXs}& \textbf{Market Capitalization /M US\$}\\ \hline
		\textsc Ethereum&EVM bytecode&23,102,544&8,468 \\ \hline
		\textsc RSK&Solidity&Unknown&N/A \\ \hline
		\textsc Counterparty&EVM bytecode&12,170,386&15\\ \hline
		\textsc Stellar&Transaction chains&Unknown&139 \\ \hline	
		\textsc Monax&EVM bytecode&Unknown&N/A \\ \hline
		\textsc Lisk&JavaScript&Unknown&71 \\ \hline
	\end{tabular}
	\vspace{-2ex}
\end{table}

In blockchain 2.0 stage, smart contract is introduced so that developers can create various applications through smart contracts. A smart contract can be considered as a lightweight dAPP (decentralized application). Ethereum is a typical system of blockchain 2.0.  Each Ethereum node runs an EVM (Ethereum Virtual Machine) that executes smart contracts. Besides Ethereum, several other blockchain systems also support smart contracts, whose information is listed in Table~\ref{smart}~\cite{bartoletti2017empirical}. In Ethereum, developers can use a variety of programming languages to develop smart contracts, such as Solidity (the recommended language), Serpent, and LLL. Since these languages are Turing-complete, smart contracts can achieve rich functions. Fig.\ref{dapp} shows the process of smart contracts' development, deployment and interaction. Each deployed smart contract corresponds to a unique address, through which users can interact with the smart contract through transactions by different clients (e.g., Parity, Geth, etc.). Since smart contracts can call each other through messages, developers can develop more feature-rich dAPPs based on available smart contracts. Compared with the traditional application, a dAPP has the following characteristics and advantages~\cite{22}:

(1). Autonomy. dAPPs are developed on the basis of smart contracts, and smart contracts are deployed and run on the blockchain. Hence, dAPPs can run autonomically without the need of any third party's assistance and participation.

(2). Stable. The bytecodes of smart contracts are stored in the state tree of blockchain. Each full node saves the information of all blocks and stateDB, including the bytecodes of smart contracts. Hence, the failure of some nodes will not affect its operation. This mechanism ensures that dAPPs can run stably.
	
(3). Traceable. Since the invocation information of smart contracts is stored in the blockchain as transactions, all the behaviors of dAPPs are recorded and traceable.
	
(4). Secure. The public key cryptography and the blockchain consensus mechanism can ensure the security and correct operations of smart contracts, so as to maximize the security of dAPPs.

\section{Risks to Blockchain}
\label{sec:risks}
\begin{table}[ht!]
	\centering
	\vspace*{-1ex}
	\scriptsize
	\caption{Taxonomy of blockchain's risks}
	\vspace{1ex}
	\label{tab_risks}
	\begin{tabular}{|c|c|c|c|}
		\hline
		\textbf{Number}&\textbf{Risk}&\textbf{Cause}&\textbf{Range of Influence}\\ \hline
		\textsc 3.1.1&51\% vulnerability&Consensus mechanism&\multirow{5}{*}{Blockchain1.0, 2.0}\\ \cline{1-3}
		\textsc 3.1.2&Private key security&Public-key encryption scheme&\\ \cline{1-3}
		\textsc 3.1.3&Criminal activity&Cryptocurrency application&\\ \cline{1-3}
		\textsc 3.1.4&Double spending&Transaction verification mechanism&\\ \cline{1-3}
		\textsc 3.1.5&Transaction privacy leakage&Transaction design flaw&\\ \hline
		\textsc 3.2.1&Criminal smart contracts&Smart contract application&\multirow{4}{*}{Blockchain2.0}\\ \cline{1-3}
		\textsc 3.2.2&Vulnerabilities in smart contract&Program design flaw&\\ \cline{1-3}
		\textsc 3.2.3&Under-optimized smart contract&Program writing flaw&\\ \cline{1-3}	
		\textsc 3.2.4&Under-priced operations&EVM design flaw&\\ \hline
	\end{tabular}
	\vspace{-2ex}
\end{table}
We divide the common blockchain risks into nine categories, as shown in Table~\ref{tab_risks}, and detail the causes and possible consequence of each risk. The risks described in Section 3.1 exist in blockchain 1.0 and 2.0, and their causes are mostly related to the blockchain operation mechanism. By contrast, the risks introduced in Section 3.2 are unique to blockchain 2.0, and are usually resulted from the development, deployment, and execution of smart contracts.

\subsection{Common Risks to Blockchain 1.0 and 2.0}
\subsubsection{51\% Vulnerability}
\label{sec:51}

The blockchain relies on the distributed consensus mechanism to establish mutual trust. However, the consensus mechanism itself has 51\% vulnerability, which can be exploited by attackers to control the entire blockchain. More precisely, in PoW-based blockchains, if a single miner's hashing power accounts for more than 50\% of the total hashing power of the entire blockchain, then the 51\% attack may be launched. Hence, the mining power concentrating in a few mining pools may result in the fears of an inadvertent situation, such as a single pool controls more than half of all computing power. In Jan. 2014, after the mining pool \texttt{ghash.io} reached 42\% of the total Bitcoin computing power, a number of miners voluntarily dropped out of the pool, and \texttt{ghash.io} issued a press statement to reassure the Bitcoin community that it would avoid reaching the 51\% threshold~\cite{18}. In PoS-based blockchains, 51\% attack may also occur if the number of coins owned by a single miner is more than 50\% of the total blockchain. By launching the 51\% attack, an attacker can arbitrarily manipulate and modify the blockchain information. Specifically, an attacker can exploit this vulnerability to conduct the following attacks~\cite{9}:

(1). Reverse transactions and initiate double spending attack (the same coins are spent multiple times).

(2). Exclude and modify the ordering of transactions.

(3). Hamper normal mining operations of other miners.

(4). Impede the confirmation operation of normal transactions.

\subsubsection{Private Key Security}

When using blockchain, the user's private key is regarded as the identity and security credential, which is generated and maintained by the user instead of third-party agencies. For example, when creating a cold storage wallet in Bitcoin blockchain, the user must import his/her private key. Hartwig et al.~\cite{mayerecdsa} discover a vulnerability in ECDSA (Elliptic Curve Digital Signature Algorithm) scheme, through which an attacker can recover the user's private key because it does not generate enough randomness during the signature process.

Once the user's private key is lost, it will not be able to be recovered. If the private key is stolen by criminals, the user's blockchain account will face the risk of being tampered by others. Since the blockchain is not dependent on any centralized third-party trusted institutions, if the user's private key is stolen, it is difficult to track the criminal's behaviors and recover the modified blockchain information.

\subsubsection{Criminal Activity}
\label{sec:criminal}
\begin{table}[ht!]
	\centering
	\vspace*{-1ex}
	\scriptsize
	\caption{Top 10 categories of items available in \texttt{Silk Road}}
	\vspace{1ex}
	\label{top20}
	\begin{tabular}{|c|c|c|c|}
		\hline
		\textbf{Number}&\textbf{Category}&\textbf{Items} & \textbf{Percentage}\\ \hline
		\textsc 1&Weed&3338&13.7\% \\ \hline
		\textsc 2&Drugs&2194&9.0\%\\ \hline
		\textsc 3&Prescription&1784&7.3\% \\ \hline
		\textsc 4&Benzos&1193&4.9\% \\ \hline
		\textsc 5&Books&955&3.9\% \\ \hline
		\textsc 6&Cannabis&877&3.6\% \\ \hline
		\textsc 7&Hash&820&3.4\% \\ \hline
		\textsc 8&Cocaine&630&2.6\% \\ \hline
		\textsc 9&Pills&473&1.9\% \\ \hline
		\textsc 10&Blotter (LSD)&440&1.8\% \\ \hline
		
	\end{tabular}
	\vspace{-2ex}
\end{table}
Bitcoin users can have multiple Bitcoin addresses, and the address has no relationship with their real life identity. Therefore, Bitcoin has been used in illegal activities. Through some third-party trading platforms that support Bitcoin, users can buy or sell any product. Since this process is anonymous, it is hard to track user behaviors, let alone subject to legal sanctions. Some frequent criminal activities with Bitcoin include:

(1). Ransomware. The criminals often use ransomware for money extortion, and employ Bitcoin as trading currency. In July 2014, a ransomware named \texttt{CTB-Locker}~\cite{27} spread around the world by disguising itself as mail attachments. If the user clicks the attachment, the ransomware will run in the background of the system and encrypt about 114 types of each file~\cite{ctblocker}. The victim has to pay the attacker a certain amount of Bitcoin within 96 hours. Otherwise, the encrypted files will not be restored. In May 2017, another ransomware \texttt{WannaCry} (also named as \texttt{WannaCrypt})~\cite{28} infected about 230,000 victims across 150 countries in two days. It exploited a vulnerability in Windows system to spread, and encrypted users' files to ask for Bitcoin ransom.

(2). Underground market. Bitcoin is often used as the currency in the underground market. For example, \texttt{Silk Road} is an anonymous, international online marketplace that operates as a Tor hidden service and uses Bitcoin as its exchange currency~\cite{christin2013traveling}. The top 10 categories of items available in \texttt{Silk Road} are listed in Table~\ref{top20}~\cite{christin2013traveling}. Most of the items sold in \texttt{Silk Road} are drugs, or some other controlled items in the real world. Since international transactions account for a significant proportion in \texttt{Silk Road}, Bitcoin makes the transaction in the underground market more convenient, which will cause harm to the social security.

(3). Money laundering. Since Bitcoin has the features like anonymity and network virtual payment and has been adopted by many countries, compared with other currencies, Bitcoin carries the lowest risk of being used for money laundering~\cite{treasury2015uk}. Cody et al. propose \texttt{Dark Wallet}~\cite{29},  a Bitcoin application that can make Bitcoin transaction completely stealth and private. \texttt{Dark Wallet} can encrypt transaction information and mix the user's valid coins with chaff coins, and hence it can make money laundering much easier.

\subsubsection{Double Spending}
\begin{figure}[ht]
	\centering
	\includegraphics[width=4in]{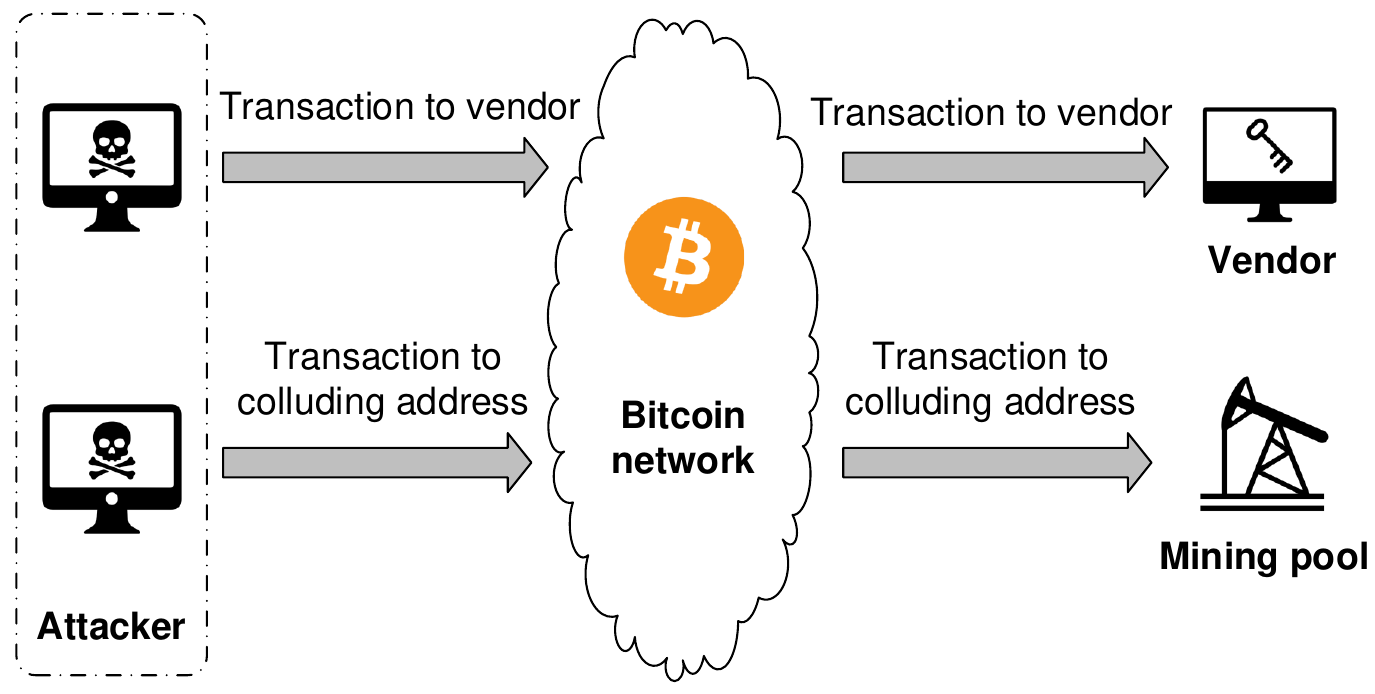}
	\caption{Double spending attack model against fast payment in Bitcoin}
	\label{double}
\end{figure}
Although the consensus mechanism of blockchain can validate transactions, it is still impossible to avoid double spending~\cite{karame2015misbehavior}. Double spending refers to that a consumer uses the same cryptocurrency multiple times for transactions. For example, an attacker could leverage race attack for double spending. This kind of attack is relatively easy to implement in PoW-based blockchains, because the attacker can exploit the intermediate time between two transactions' initiation and confirmation to quickly launch an attack. Before the second transaction is mined to be invalid, the attacker has already got the first transaction's output, resulting in double spending.

Ghassan et al.~\cite{karame2012double} conduct an analysis of double spending against fast payment in Bitcoin, and propose an attack model, as shown in Fig.\ref{double}. Assuming that an attacker knows the vendor's address before the attack, to perform double spending, the attacker will send two transactions, $TX_{v}$ and $TX_{a}$ and choose the same BTCs (cryptocurrency in Bitcoin) as inputs for $TX_{v}$ and $TX_{a}$. $TX_{v}$'s recipient address is set to the targeted vendor's address, and $TX_{a}$'s recipient address is set to the colluding address controlled by the attacker. If the following three conditions are met, double spending will be successful: (1) $TX_{v}$ is added to the wallet of the targeted vendor; (2) $TX_{a}$ is mined as valid into the blockchain; (3) The attacker gets $TX_{v}$'s output before the vendor detects misbehavior. If the attack is successful, $TX_{v}$ will eventually be verified as an invalid transaction, and BTCs are really spent by $TX_{a}$. The attacker has received $TX_{v}$'s output, which is the vendor's normal service. Since $TX_{a}$'s recipient address is controlled by the attacker, these BTCs are still owned by herself. In this double spending model, the attacker enjoys the service without paying any BTC.

\subsubsection{Transaction Privacy Leakage}
\label{sec:privacy}
Since the users' behaviors in the blockchain are traceable, the blockchain systems take measures to protect the transaction privacy of users. In the Bitcoin and Zcash, they use one-time accounts to store the received cryptocurrency. Moreover, the user needs to assign a private key to each transaction. In this way, the attacker cannot infer whether the cryptocurrency in different transactions is received by the same user. In Monero, users can include some chaff coins (called ``mixins'') when they initiate a transaction so that the attacker cannot infer the linkage of actual coins spent by the transaction.

\begin{table}[ht!]
	\centering
	\vspace*{-1ex}
	\scriptsize
	\caption{Linkability analysis of Monera transaction inputs with mixins}
	\vspace{1ex}
	\label{leak}
	\begin{tabular}{|c|c|c|c|}
		\hline
		\textbf{}&\textbf{Not deducible} & \textbf{Deducible}& \textbf{In total}\\ \hline
		\textsc Using newest TXO&15.07\%&4.60\%&19.67\% \\ \hline
		\textsc Not using newest TXO&22.61\%&57.72\%&80.33\%\\ \hline
		\textsc In total&37.68\%&62.32\%&100\% \\ \hline	
	\end{tabular}
	\vspace{-2ex}
\end{table}
Unfortunately, the privacy protection measures in blockchain are not very robust. Andrews et al.~\cite{miller2017empirical} empirically evaluate two linkability weaknesses in Monero's mixin sampling strategy, and discover that 66.09\% of all transactions do not contain any mixins. 0-mixin transaction will lead to the privacy leakage of its sender. Since users may use the outputs of 0-mixin transaction as mixins, these mixins will be deducible. Moreover, they study the sampling method of mixins and find that the selection of mixins is not really random. Newer TXOs (transaction outputs) tend to be used more frequently. They further discover that 62.32\% of transaction inputs with mixins are deducible, as shown in Table~\ref{leak}~\cite{miller2017empirical}. By exploiting these weaknesses in Monero, they can infer the actual transaction inputs with 80\% accuracy.

\subsection{Specific Risks to Blockchain 2.0}

\subsubsection{Criminal Smart Contracts}
\begin{figure}[ht]
	\centering
	\includegraphics[width=4in]{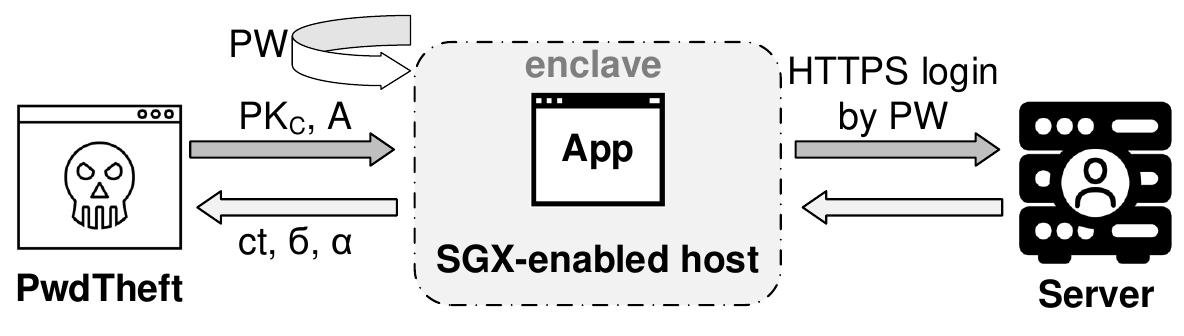}
	\caption{Execution procedure of \texttt{PwdTheft} using SGX-enabled platform}
	\label{PwdTheft}
\end{figure}

Criminals can leverage smart contracts for a variety of malicious activities, which may pose a threat to our daily life. CSCs (Criminal Smart Contracts) can facilitate the leakage of confidential information, theft of cryptographic keys, and various real-world crimes (e.g., murder, arson, terrorism, etc.)~\cite{juels2015ring}. Juels et al. propose an example of password theft CSC \texttt{PwdTheft}, whose process is shown in Fig.\ref{PwdTheft}~\cite{juels2015ring}. \texttt{PwdTheft} can be exploited for a fair exchange between contractor \texttt{C} and perpetrator \texttt{P}. \texttt{C} will pay a reward to \texttt{P} if and only if \texttt{P} gives a valid password to \texttt{C}. The entire transaction process can be done without any third party trusted agencies involved. Since the smart contract deployed in blockchain cannot access network directly~\cite{zhang2016town}, in the actual work process of \texttt{PwdTheft}, it is combined with trusted hardware technology, such as Intel SGX (Software Guard eXtension), to prove the validity of the password through HTTPS (Hypertext Transfer Protocol Secure). SGX will create a trusted execution environment named \texttt{enclave}, which can protect the application from being attacked by others. Any privileged or unprivileged software cannot access the runtime environment of \texttt{enclave}. Furthermore, SGX can produce \texttt{quote}, a digitally signed attestation. \texttt{Quote} can get the hash value of the application run in \texttt{enclave} environment. Meanwhile, \texttt{quote} can access the relevant data during runtime of the application. The whole password exchange process is divided into three steps:

(1). \texttt{PwdTheft} provides $(pk_{C}, A)$, $pk_{C}$ is the public key of \texttt{C}, and \texttt{A} is the target account for stealing.

(2). The application that runs in SGX, using the \texttt{PW} provided by \texttt{P}, logs on to the server account \texttt{A} by establishing an HTTPS connection.

(3). If the preceding steps are successful, the data \texttt{ct}, $\sigma$ and $\alpha$ will be transmitted to \texttt{PwdTheft}. \texttt{ct} = $enc_{pk_{C}}[PW]$ and $\sigma$ = $Sig_{sk_{app}} [ct]$. $sk_{app}$ is the signature private key of the application. $\alpha$ is a \texttt{quote} that runs on \texttt{P}'s SGX-enabled host.

After \texttt{PwdTheft} receives \texttt{ct}, $\sigma$ and $\alpha$, \texttt{C} can decrypt them to verify the data, and then decide whether a reward should be paid to \texttt{P}. In this process, in order to prevent \texttt{P} from changing the password maliciously after the data transmission to \texttt{PwdTheft}, they can add a timestamp in the data. In addition, \texttt{PwdTheft} can be easily extended for conducting other malicious activities. For example, criminals can leverage CSCs to make 0-day vulnerability transactions, which are critical cyber-weaponry~\cite{juels2015ring}.

\subsubsection{Vulnerabilities in Smart Contract}
\label{sec:vul}
\begin{table}[ht!]
	\centering
	\vspace*{-1ex}
	\scriptsize
	\caption{Taxonomy of vulnerabilities in smart contract}
	\vspace{1ex}
	\label{vul}
	\begin{tabular}{|c|c|c|c|}
		\hline
		\textbf{Number}&\textbf{Vulnerability} & \textbf{Cause}& \textbf{Level}\\ \hline
		\textsc 1&Call to the unknown&The called function does not exist&\multirow{6}{*}{Contract source code} \\ \cline{1-3}
		\textsc 2&Out-of-gas send&Fallback of the callee is executed& \\ \cline{1-3}
		\textsc 3&Exception disorder&Irregularity in exception handling& \\ \cline{1-3}	
		\textsc 4&Type casts&Type-check error in contract execution& \\ \cline{1-3}
		\textsc 5&Reentrancy vulnerability&Function is re-entered before termination& \\ \cline{1-3}
		\textsc 6&Field disclosure&Private value is published by the miner& \\ \hline
		\textsc 7&Immutable bug&Alter a contract after deployment&\multirow{3}{*}{EVM bytecode} \\ \cline{1-3}	
		\textsc 8&Ether lost&Send Ether to an orphan address& \\ \cline{1-3}
		\textsc 9&Stack overflow&The number of values in stack exceeds 1024& \\ \hline
		\textsc 10&Unpredictable state&State of the contract is changed before invoking&\multirow{3}{*}{Blockchain mechanism} \\ \cline{1-3}
		\textsc 11&Randomness bug&Seed is biased by malicious miner& \\ \cline{1-3}	
		\textsc 12&Timestamp dependence&Timestamp of block is changed by malicious miner& \\ \hline
	\end{tabular}
	\vspace{-2ex}
\end{table}
As programs running in the blockchain, smart contracts may have security vulnerabilities caused by program defects. Nicola et al.~\cite{atzeisurvey} conduct a systematic investigation of 12 types of vulnerabilities in smart contract, as shown in Table~\ref{vul}. Loi et al.~\cite{luu2016making} propose a symbolic execution tool called \textsc{Oyente} to find 4 kinds of potential security bugs. They discover that 8,833 out of 19,366  Ethereum smart contracts are vulnerable. The details of these 4 bugs are as follows:

(1). Transaction-ordering dependence. Valid transactions can change the state of Ethereum blockchain from $\sigma$ to $\sigma'$:  ${\sigma \overset{T}{\rightarrow} \sigma'}$ . In every epoch, each miner proposes their own block to update the blockchain. Since a block may contain multiple transactions, blockchain state $\sigma$ may change multiple times within an epoch. When a new block contains two transactions $T_{i}$ and $T_{j}$, which invoke the same smart contract, it may trigger this vulnerability. Because the execution of the smart contract is associated with state $\sigma$, the execution order of $T_{i}$ and $T_{j}$ affects the ultimate state. The order of transactions' execution depends entirely on miners. In this case, TOD (Transaction-Ordering Dependent) contracts are vulnerable.

(2). Timestamp dependence. In the blockchain, every block has a \texttt{timestamp}. Some smart contracts' trigger conditions depend on \texttt{timestamp}, which is set by the miner according to its local system time. If an attacker can modify it, timestamp-dependent contracts are vulnerable.

(3). Mishandled exceptions. This category of vulnerability may occur when different smart contracts are called from each other. When contract \texttt{A} calls contract \texttt{B}, if \texttt{B} runs abnormally, \texttt{B} will stop running and return \texttt{false}. In some invocations, contract \texttt{A} must explicitly check the return value to verify if the call has been executed properly. If \texttt{A} does not correctly check the exception information, it may be vulnerable.

(4). Reentrancy vulnerability. During the invocation of the smart contract, the actual state of the contract account is changed after the call is completed. An attacker can use the intermediate state to conduct repeated calls to the smart contract. If the invoked contract involves Ether transaction, it may result in illegal Ether stealing.

\subsubsection{Under-Optimized Smart Contract}
\begin{table}[ht!]
	\centering
	\vspace*{-1ex}
	\scriptsize
	\caption{Taxonomy of under-optimized patterns in smart contract}
	\vspace{1ex}
	\label{under}
	\begin{tabular}{|c|c|c|}
		\hline
		\textbf{Number}&\textbf{Under-optimized pattern} & \textbf{Category}\\ \hline
		\textsc 1&Dead code&\multirow{2}{*}{Useless-code related patterns} \\ \cline{1-2}
		\textsc 2&Opaque predicate& \\ \hline
		\textsc 3&Expensive operations&\multirow{5}{*}{Loop-related patterns} \\ \cline{1-2}	
		\textsc 4&Constant outcome& \\ \cline{1-2}
		\textsc 5&Loop fusion& \\ \cline{1-2}
		\textsc 6&Repeated computations& \\ \cline{1-2}
		\textsc 7&Comparison with unilateral outcome& \\ \hline
	\end{tabular}
	\vspace{-2ex}
\end{table}
When a user interacts with a smart contract deployed in Ethereum, a certain amount of gas is charged. Gas can be exchanged with Ether, which is the cryptocurrency in Ethereum. Unfortunately, some smart contracts' development and deployment are not adequately optimized. Chen et al.~\cite{chen2017under} identify 7 gas-costly patterns and group them into 2 categories (as shown in Table~\ref{under}): useless-code related patterns, and loop-related patterns. They propose a tool named \textsc{Gasper}, which can automatically discover 3 gas-costly patterns in smart contracts: dead code, opaque predicate, and expensive operations in a loop. Leveraging \textsc{Gasper}, they find that more than 80\% smart contracts deployed in Ethereum (4,240 real smart contracts) have at least one of these 3 patterns. The details are as follows:

(1). Dead code. It means that some operations in a smart contract will never be executed, but they will still be deployed into the blockchain. Since in the smart contract deployment process the consumption of gas is related to bytecode size, the dead code will cause additional gas consumption.

(2). Opaque predicate. For some statements in a smart contract, their execution results are always the same and will not affect other statements and the smart contract. The presence of the opaque predicate causes the EVM to execute useless operations, thereby consuming additional gas.

(3). Expensive operations in a loop. It refers to some expensive operations within a loop, which can be moved outside the loop to save gas consumption.

\subsubsection{Under-Priced Operations}
As mentioned earlier, each operation is set to a specific gas value in Ethereum, which can be queried in the yellow paper~\cite{26}. Ethereum sets the gas value based on the execution time, bandwidth, memory occupancy and other parameters. In general, the gas value is proportional to the computing resources consumed by the operation. However, it is difficult to accurately measure the consumption of computing resources of an individual operation, and therefore some gas values are not set properly. For example, some IO-heavy operations' gas values are set too low, and hence these operations can be executed in quantity in one transaction. In this way, an attacker can initiate a DoS (Denial of Service) attack on Ethereum.

\begin{table}[ht!]
	\centering
	\vspace*{-1ex}
	\scriptsize
	\caption{Gas table modification in EIP150}
	\vspace{1ex}
	\label{EIP150}
	\begin{tabular}{|c|c|c|c|}
		\hline
		\textbf{Number}&\textbf{Operation}&\textbf{Old value} & \textbf{EIP150 value}\\ \hline
		\textsc 1&EXTCODESIZE&20&700 \\ \hline
		\textsc 2&EXTCODECOPY&20&700\\ \hline
		\textsc 3&BALANCE&20&400 \\ \hline
		\textsc 4&SLOAD&50&200 \\ \hline
		\textsc 5&CALL&40&700 \\ \hline
		\textsc 6&SUICIDE (does not create account)&0&5,000 \\ \hline	
		\textsc 7&SUICIDE (creates an account)&0&25,000 \\ \hline
	\end{tabular}
	\vspace{-2ex}
\end{table}

Actually, attackers have exploited the under-priced operation \texttt{EXTCODESIZE} to attack Ethereum~\cite{23}. When \texttt{EXTCODESIZE} is executed, it needs to read state information and then the node will read hard disk. Since the gas value of \texttt{EXTCODESIZE} is only 20, the attacker can call it more than 50,000 times in one transaction. This will cause the user to consume a lot of computing resources, and block synchronization will be significantly slower compared with the normal situation. As another example, some attackers exploited the under-priced operation \texttt{SUICIDE} to launch DoS attacks~\cite{24}. They exploited \texttt{SUICIDE} to create about 19 million empty accounts, which need to be stored in the state tree. This attack caused a waste of hard disk resources. At the same time, the node information synchronization and transaction processing speed are significantly decreased.

In order to solve the security problem caused by under-priced operations, the gas values of 7 IO-heavy operations are modified in EIP (Ethereum Improvement Proposal) 150~\cite{25}, as shown in Table~\ref{EIP150}. Note that EIP150 has already been implemented in the Ethereum public chain by hard fork, and the new gas table parameters are used after No.2463000 block.

\section{Attack Cases}
\label{sec:attackcases}
In this section, we survey real attacks on blockchain systems, and analyze the vulnerabilities exploited in these attacks.

\subsection{Selfish Mining Attack}
The selfish mining attack is conducted by attackers (i.e., selfish miners) for the purpose of obtaining undue rewards or wasting the computing power of honest miners \cite{SolatP16}. The attacker holds discovered blocks privately and then attempts to fork a private chain \cite{EyalS14}. Afterwards, selfish miners would mine on this private chain, and try to maintain a longer private branch than the public branch because they privately hold more newly discovered blocks. In the meanwhile, honest miners continue mining on the public chain. New blocks mined by the attacker would be revealed when the public branch approaches the length of private branch, such that the honest miners end up wasting computing power and gaining no reward, because selfish miners publish their new blocks just before honest miners. As a result, the selfish miners gain a competitive advantage, and honest miners would be incentivized to join the branch maintained by selfish miners. Through a further consolidation of mining power into the attacker's favor, this attack undermines the decentralization nature of blockchain.

Ittay et al.~\cite{EyalS14} propose an attack strategy named \textsc{Selfish-Mine}, which can force the honest miners to perform wasted computations on the stale public branch. In the initial circumstance of \textsc{Selfish-Mine}, the length of the public chain and private chain are the same. The \textsc{Selfish-Mine} involves the following three scenarios:

(1). The public chain is longer than the private chain. Since the computing power of selfish miners may be less than that of the honest miners, selfish miners will update the private chain according to the public chain, and in this scenario, selfish miners cannot gain any reward.

(2). Selfish miners and honest miners almost simultaneously find the first new block. In this scenario, selfish miners will publish the newly discovered block, and there will be two concurrently forks of the same length. Honest miners will mine in either of the two branches, while selfish miners will continue to mine on the private chain. If selfish miners firstly find the second new block, they will publish this block immediately. At this point, selfish miners will gain two blocks' rewards at the same time. Because the private chain is longer than the public chain, the private chain will be the ultimate valid branch. If honest miners firstly find the second new block and this block is written to the private chain, selfish miners will gain the first new block' rewards, and honest miners will gain the second new block' rewards. Otherwise, if this block is written to the public block, honest miners will gain these two new blocks' rewards, and selfish miners will not gain any reward.

(3). After selfish miners find the first new block, they also find the second new block. In this scenario, selfish miners will hold these two new blocks privately, and they continue to mine new blocks on the private chain. When the first new block is found by honest miners, selfish miners will publish its own first new block. When honest miners find the second new block, the selfish miners will immediately publish its own second new block. Then selfish miners will follow this response in turn, until the length of the public chain is only 1 greater than the private chain, after which the selfish miners will publish its last new block before honest miners find this block. At this point, the private chain will be considered valid, and consequently selfish miners will gain the rewards of all new blocks.

\subsection{DAO Attack}

\begin{table}[ht!]
	\centering
	\vspace*{-1ex}
	\scriptsize
	\caption{Some other attacks that exploit smart contracts' vulnerabilities}
	\vspace{1ex}
	\label{vulcase}
	\begin{tabular}{|c|c|c|}
		\hline
		\textbf{Number}&\textbf{Attack case} & \textbf{Related vulnerabilities}\\ \hline
		\textsc 1&King of the Ether throne&\tabincell{c}{Out-of-gas send\\Exception disorder} \\ \hline
		\textsc 2&Multi-player games&Field disclosure \\ \hline
		\textsc 3&Rubixi attack&Immutable bug \\ \hline	
		\textsc 4&GovernMental attack&\tabincell{c}{Immutable bug\\Stack overflow\\Unpredictable state\\Timestamp dependence} \\ \hline
		\textsc 5&Dynamic libraries attack&Unpredictable state \\ \hline
	\end{tabular}
	\vspace{-2ex}
\end{table}
The \texttt{DAO} is a smart contract deployed in Ethereum on 28th May of 2016, which implements a crowd-funding platform. The \texttt{DAO} contract was attacked only after it has been deployed for 20 days. Before the attack happened, \texttt{DAO} has already raised 150 million US\$, which is the biggest crowdfund ever. The attacker stole about 60 million US\$.

The attacker exploited the reentrancy vulnerability in this case. Firstly, the attacker publishes a malicious smart contract, which includes a  \texttt{withdraw()} function \texttt{call} to \texttt{DAO} in its callback function. The \texttt{withdraw()} will send Ether to the callee, which is also in the form of \texttt{call}. Therefore, it will invoke the callback function of the malicious smart contract again. In this way, the attacker is able to steal all the Ether from \texttt{DAO}. There are some other cases that exploit smart contracts' vulnerabilities (described in Section \ref{sec:vul}), which are listed in Table~\ref{vulcase}~\cite{atzeisurvey}.

\subsection{BGP Hijacking Attack}

BGP (Border Gateway Protocol) is a de-facto routing protocol and regulates how IP packets are forwarded to their destination. To intercept the network traffic of blockchain, attackers either leverage or manipulate BGP routing. BGP hijacking typically requires the control of network operators, which could potentially be exploited to delay network messages. Maria et al.~\cite{ApostolakiZV16} comprehensively analyze the impact of routing attacks, including  both node-level and network-level attacks, on Bitcoin, and show that the number of the successfully to-be-hijacked Internet prefixes depends on the distribution of mining power. Because of the high centralization of some Bitcoin mining pools, if they are attacked by BGP hijacking, it will have a significant effect. The attackers can effectively split the Bitcoin network, or delay the speed of block propagation.

Attackers conduct BGP hijacking to intercept Bitcoin miners' connections to a mining pool server, as analyzed by Dell SecureWorks in 2014~\cite{BGPHijacking}. By rerouting traffic to a mining pool controlled by the attacker, it was possible to steal cryptocurrency from the victim. This attack collected an estimated 83,000 US\$ of cryptocurrency over a two month period.  Since the BGP security extensions are not widely deployed, network operators have to rely on monitoring systems, which would report rogue announcements, such as BGPMon~\cite{4804446}. However, even if an attack is detected, solving a hijacking still cost hours as it is a human-driven process consisting of altering configuration or disconnecting the attacker. For example, YouTube ever took about three hours to resolve a hijacking of its prefixes by a Pakistani ISP (Internet Service Provider)~\cite{Youtube}.

\subsection{Eclipse Attack}
\begin{table}[ht!]
	\centering
	\vspace*{-1ex}
	\scriptsize
	\caption{Some other attacks that may be caused by the eclipse attack}
	\vspace{1ex}
	\label{eclipse}
	\begin{tabular}{|c|c|c|}
		\hline
		\textbf{Number}&\textbf{Attack} & \textbf{Harm}\\ \hline
		\textsc 1&Engineering block races&Wasting mining power on orphan blocks \\ \hline
		\textsc 2&Splitting mining power&51\% vulnerability may be triggered \\ \hline
		\textsc 3&Selfish mining&Attacker can gain more than normal mining rewards \\ \hline	
		\textsc 4&0-confirmation double spend&\multirow{2}{*}{The vendor would not get rewards for its service} \\ \cline{1-2}
		\textsc 5&N-confirmation double spend& \\ \hline
	\end{tabular}
	\vspace{-2ex}
\end{table}
The eclipse attack allows an attacker to monopolize all of the victim's incoming and outgoing connections, which isolates the victim from the other peers in the network \cite{SinghNDW06}. Then, the attacker can filter the victim's view of the blockchain, or let the victim cost unnecessary computing power on obsolete views of the blockchain. Furthermore, the attacker is able to leverage the victim's computing power to conduct its own malicious acts. Ethan et al.~\cite{HeilmanKZG15} consider two types of eclipse attack on Bitcoin's peer-to-peer network, namely botnet attack and infrastructure attack. The botnet attack is launched by bots with diverse IP address ranges. The infrastructure attack models the threat from an ISP, company or nation-state that has contiguous IP addresses. The Bitcoin network might suffer from disruption and a victim's view of the blockchain will be filtered due to the eclipse attack. Additionally, the eclipse attack is a useful basis for other attacks, as shown in Table~\ref{eclipse}~\cite{HeilmanKZG15}.

\subsection{Liveness Attack}

\begin{figure}[ht]
	\centering
	\includegraphics[width=4in]{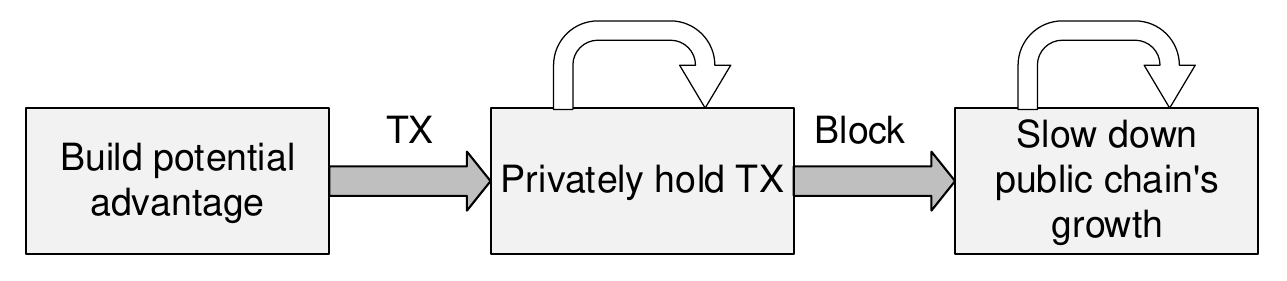}
	\caption{Overview of the liveness attack process}
	\label{Livenessattack}
\end{figure}

Aggelos et al.~\cite{KiayiasP16} propose the liveness attack, which is able to delay as much as possible the confirmation time of a target transaction. They also present two instantiations of such attack on Bitcoin and Ethereum. Liveness attack consists of three phases, namely attack preparation phase, transaction denial phase, and blockchain retarder phase (shown in Fig.\ref{Livenessattack}):

(1). Attack preparation phase. Just like selfish mining attack, an attacker builds advantage over honest miners in some way before the target transaction TX is broadcasted to the public chain. The attacker builds the private chain, which is longer than the public chain.

(2). Transaction denial phase. The attacker privately holds the block that contains TX, in order to prevent TX from being written into the public chain.

(3). Blockchain retarder phase. In the growth process of the public chain, TX will no longer be able to be privately held in a certain time. In this case, the attacker will publish the block that contains TX. In some blockchain systems, when the depth of the block that contains TX is greater than a constant, TX will be regarded valid. Therefore, the attacker will continue building private chain in order to build an advantage over the public chain. After that, the attacker will publish her privately held blocks into public chain in proper time to slow down the growth rate of public chain. The liveness attack will end when TX is verified as valid in the public chain.

\subsection{Balance Attack}
Christopher et al.~\cite{NatoliG16a} propose the balance attack against PoW-based blockchain, which allows a low-mining-power attacker to momently disrupt communications between subgroups with similar mining power. They abstract blockchain into a DAG (Directed Acyclic Graph) tree, in which DAG = $<B,P>$. $B$ are the nodes indicating blocks' information, and they are connected through directed edges $P$. After introducing a delay between correct subgroups of equivalent mining power, the attacker issues transactions in one subgroup (called ``transaction subgroup'') and mines blocks in another subgroup (called ``block subgroup''), to guarantee that the tree of block subgroup outweighs the tree of transaction subgroup. Even though the transactions are committed, the attacker is able to outweigh the tree containing this transaction and rewrite blocks with high probability.

The balance attack inherently violates the persistence of the main branch prefix and allows double spending. The attacker needs to identify the merchant-involved subgroup and create transactions to purchase goods from those merchants. Thereafter, the attacker issues transactions to this subgroup and propagates the mined blocks to the rest nodes of the group. As long as the merchant ships goods, the attacker stops delaying messages. With a high probability that the DAG tree seen by the merchant is outweighed by another tree, the attacker could successfully reissue another transaction using exactly the same coins. Balance attack proves that PoW-based blockchain is block oblivious. That is, when writing a transaction into the main chain, there is a certain probability that the attacker can override or delete the block containing this transaction. In the related experiment, the authors configure an Ethereum private chain with equivalent parameters of R3 consortium~\cite{r3}, and showed that they can successfully carry out the balance attack, which only needs to control about 5\% of total computing power.

\section{Security Enhancements}
In this section, we summarize security enhancements to blockchain systems, which can be used in the development of blockchain systems.

\label{sec:enhancement}
\subsection{SmartPool}
\begin{figure}[ht]
	\centering
	\includegraphics[width=5in]{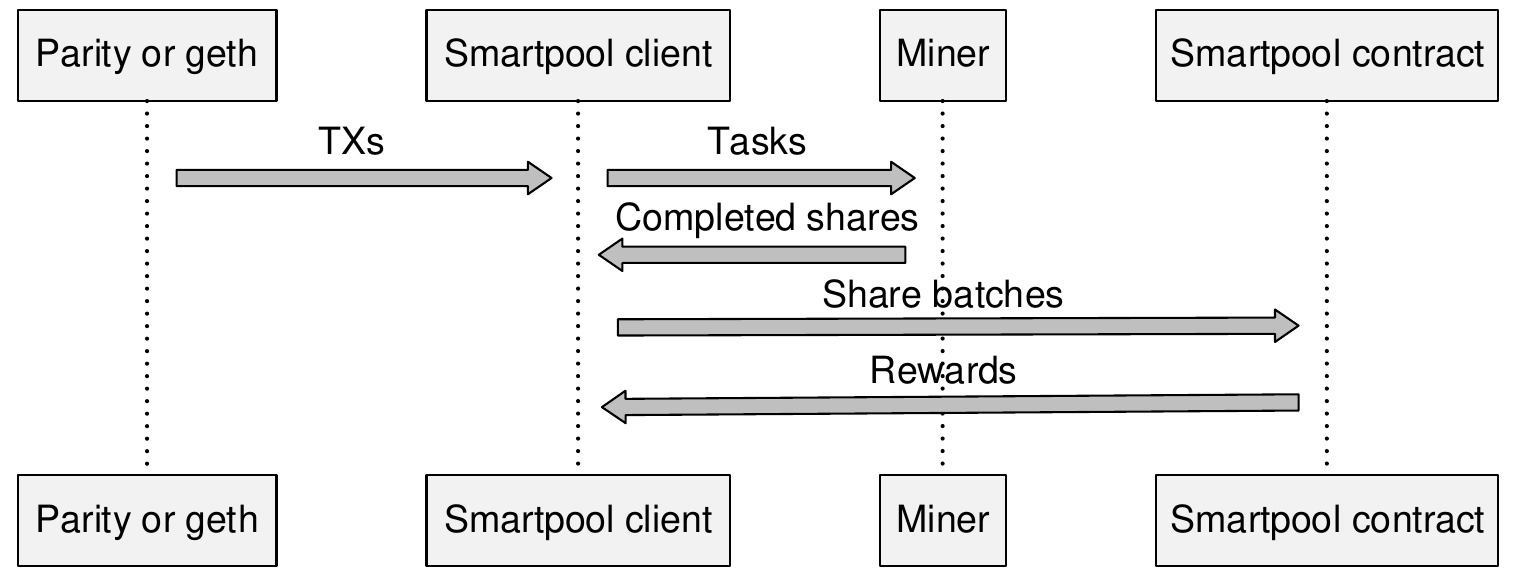}
	\caption{Overview of \textsc{SmartPool}'s execution process}
	\label{pool}
\end{figure}
As described in Section \ref{sec:51}, there already has mining pool with more than 40\% of total computing power of blockchain. This poses a serious threat to the decentralization nature, making blockchain vulnerable to several kinds of attacks. Loi et al.~\cite{luu2017smart} propose a novel mining pool system named \textsc{SmartPool}, whose workflow is shown in Fig.\ref{pool}. \textsc{SmartPool} gets the transactions from Ethereum node clients (i.e., parity~\cite{Parity} or geth~\cite{Geth}), which contain mining tasks information. Then, the miner conducts hashing computation based on the tasks and returns the completed shares to the smartpool client. When the number of the completed shares reaches to a certain amount, they will be committed to smartpool contract, which is deployed in Ethereum. The smartpool contract will verify the shares and deliver rewards to the client. Compared with the traditional P2P pool, \textsc{SmartPool} system has the following advantages:

(1). Decentralized. The core of the \textsc{SmartPool} is implemented in the form of smart contract, which is deployed in blockchain. Miners need first connect to Ethereum to mine through the client. Mining pool can rely on Ethereum's consensus mechanism to run. In this way, it ensures decentralization nature of pool miners. The mining pool state is maintained by Ethereum and no longer requires a pool operator.

(2). Efficiency. Miners can send the completed shares to the smartpool contract in batches. Furthermore, miners only need to send part of shares to be verified, not all shares. Hence, \textsc{SmartPool} is more efficient than the P2P pool.

(3). Secure. \textsc{SmartPool} leverages a novel data structure, which can prevent the attacker from resubmitting shares in different batches. Furthermore, the verification method of \textsc{SmartPool} can guarantee that honest miners will gain expected rewards even there exist malicious miners in the pool.

\subsection{Quantitative Framework}
\begin{figure}[ht]
	\centering
	\includegraphics[width=4in]{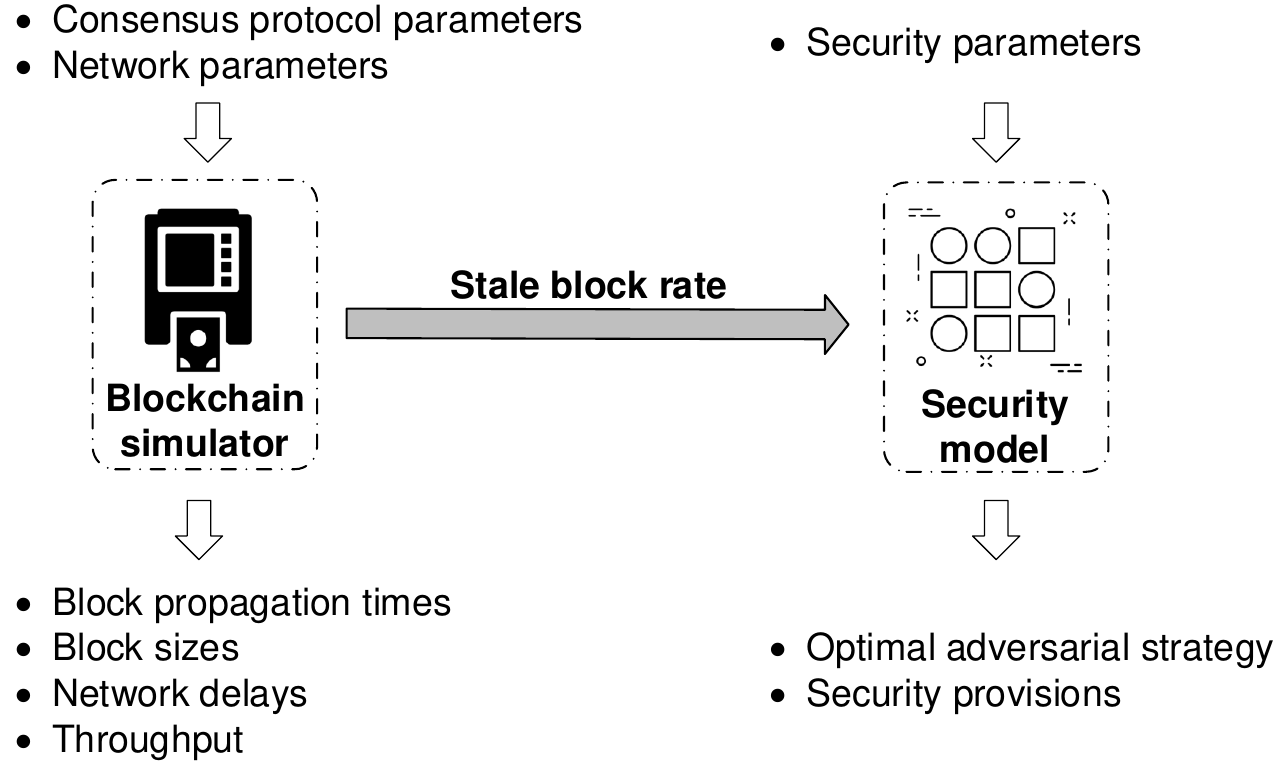}
	\caption{Components of quantitative framework}
	\label{Quantitativeframework}
\end{figure}
There exist tradeoffs between blockchain's performance and security. Arthur et al.~\cite{GervaisKWGRC16} propose a quantitative framework, which is leveraged to analyze PoW-based blockchain's execution performance and security provisions. As shown in Fig.\ref{Quantitativeframework}, the framework has two components: blockchain stimulator and security model. The stimulator mimics blockchain's execution, whose inputs are parameters of consensus protocol and network. Through the simulator's analysis, it can gain performance statistics of the target blockchain, including block propagation times, block sizes, network delays, stale block rate, throughput, etc. The stale block refers to a block that is mined but not written to the public chain. The throughput is the number of transactions that the blockchain can handle per second. Stale block rate will be passed as a parameter to the security model component, which is based on MDP (Markov Decision Processes) for defeating double spending and selfish mining attacks. The framework eventually outputs optimal adversarial strategy against attacks, and facilitates building security provisions for the blockchain.

\subsection{\textsc{Oyente}}
Loi et al.~\cite{luu2016making} propose \textsc{Oyente} to detect bugs in Ethereum smart contracts. \textsc{Oyente} leverages symbolic execution to analyze the bytecode of smart contracts and it follows the execution model of EVM. Since Ethereum stores the bytecode of smart contracts in its blockchain, \textsc{Oyente} can be used to detect bugs in deployed contracts.

\begin{figure}[ht]
	\centering
	\includegraphics[width=4in]{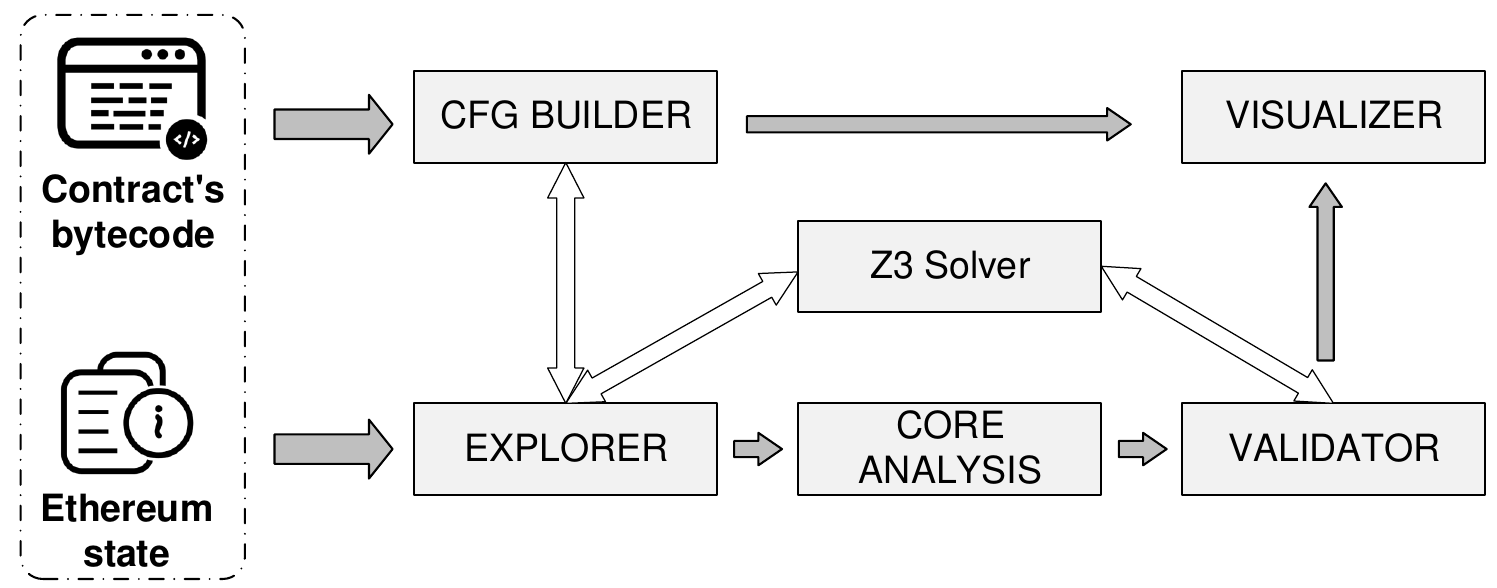}
	\caption{Overview of \textsc{Oyente}'s architecture design and execution process}
	\label{oyente}
\end{figure}
Fig.\ref{oyente} shows \textsc{Oyente}'s architecture and execution process. It takes the smart contract's bytecode and Ethereum global state as inputs. Firstly, based on the bytecode, \texttt{CFG BUILDER} will statically build CFG (Control Flow Graph) of smart contract. Then, according to Ethereum state and CFG information, \texttt{EXPLORER} conducts simulated execution of smart contract leveraging static symbolic execution. In this process, CFG will be further enriched and improved because some jump targets are not constants; instead, they should be computed during symbolic execution. The \texttt{CORE ANALYSIS} module uses the related analysis algorithms to detect four different vulnerabilities (described in Section \ref{sec:vul}). The \texttt{VALIDATOR} module validates the detected vulnerabilities and vulnerable paths. Confirmed vulnerability and CFG information will finally be output to the \texttt{VISUALIZER} module, which can be employed by users to carry out debugging and program analysis. Currently, \textsc{Oyente} is open source for public use \cite{Oyente}.

\subsection{Hawk}
As described in Section \ref{sec:privacy}, privacy leakage is a serious threat to blockchain. In the era of blockchain 2.0, not only transactions but also contract-related information are public, such as contract's bytecode, invoking parameters, etc.

Ahmed et al.~\cite{kosba2016hawk} propose \textsc{Hawk}, a novel framework for developing privacy-preserving smart contracts. Leveraging \textsc{Hawk}, developers can write private smart contracts, and it is not necessary for them to use any code encryption or obfuscation techniques. Furthermore, the financial transaction's information will not be explicitly stored in blockchain. When programmers develop \textsc{Hawk} contract, the contract can be divided into two parts: private portion, and public portion. The private data and financial function related codes can be written into the private portion, and codes that do not involve private information can be written into the public portion. The \textsc{Hawk} contract is compiled into three pieces. (1). The program that will be executed in all virtual machines of nodes, just like smart contracts in Ethereum. (2). The program that will only be executed by the users of smart contracts. (3). The program that will be executed by the manager, which is a special trustworthy party in \textsc{Hawk}. The \textsc{Hawk} manager is executed in Intel SGX enclave (described in Section \ref{sec:criminal}), and it can see the privacy information of the contract but will not disclose it. \textsc{Hawk} can not only protect privacy against the public, but also protect the privacy between different \textsc{Hawk} contracts. If the manager aborts the protocol of \textsc{Hawk}, it will be  automatically financially penalized, and the users will gain compensation. Overall, \textsc{Hawk} can largely protect the privacy of users when they are using blockchains.

\subsection{Town Crier}

\begin{figure}[ht]
	\centering
	\includegraphics[width=4.5in]{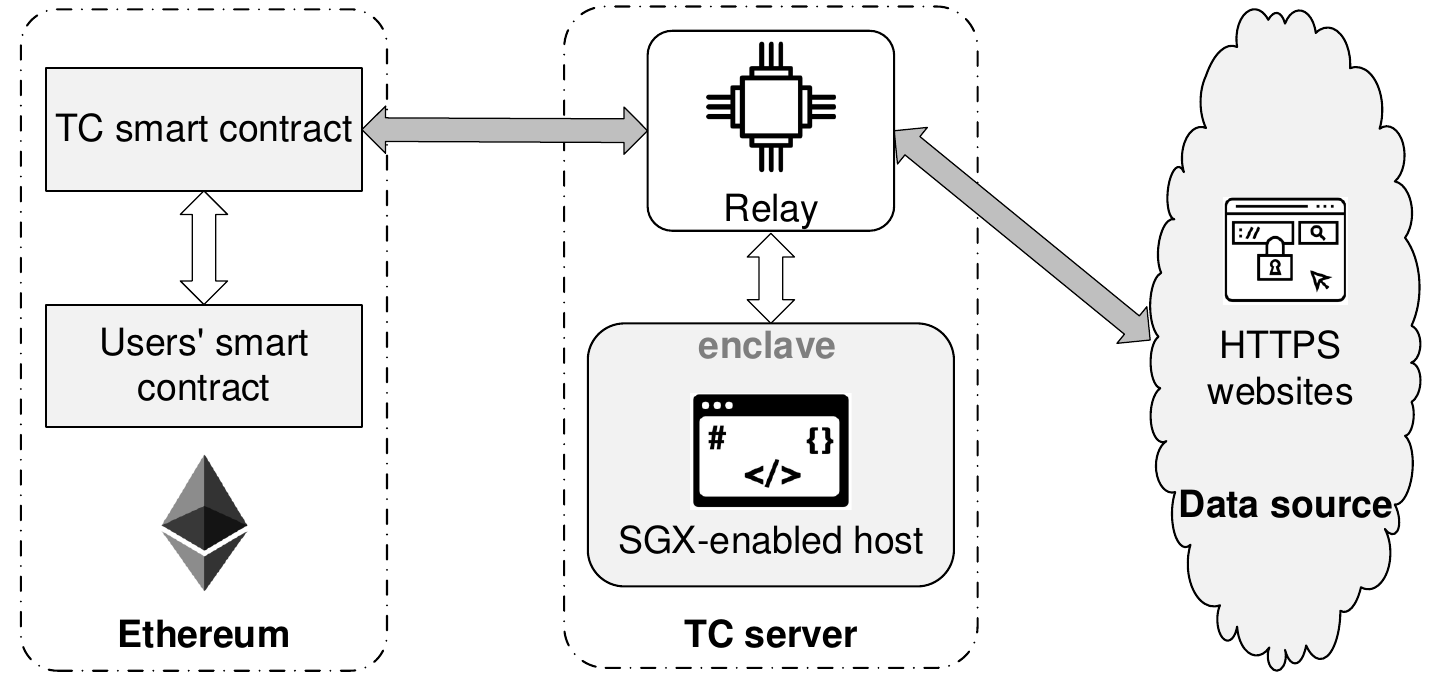}
	\caption{Basic architecture of \textsc{Town Crier} system}
	\label{tc}
\end{figure}
Smart contract often needs to interact with off-chain (i.e., external) data source. Zhang et al.~\cite{zhang2016town} propose \textsc{TC} (\textsc{Town Crier}), which is an authenticated data feed system for this data interaction process. Since the smart contract deployed in blockchain cannot access network directly, they cannot get data through HTTPS. \textsc{TC} exactly acts as a bridge between HTTPS-enabled data source and smart contracts. The basic architecture of \textsc{TC} is shown in Fig.\ref{tc}. \textsc{TC} contract is the front end of the \textsc{TC} system, which acts as API between users' contracts and \textsc{TC} server. The core program of \textsc{TC} is running in Intel SGX enclave (described in Section \ref{sec:criminal}). The main function of the \textsc{TC} server is to obtain the data requests from users' contracts, and obtain the data from target HTTPS-enabled websites. Finally, the \textsc{TC} server will return a datagram to the users' contracts in the form of digitally signed blockchain messages.

\textsc{TC} can largely protect the security of the data requesting process. The core modules of \textsc{TC} are respectively running on decentralized Ethereum, SGX-enabled \texttt{enclave}, and HTTPS-enabled website. Furthermore, the \texttt{enclave} disables the function of network connection to maximize its security. \texttt{Relay} module is designed as a network communication hub for smart contracts, SGX \texttt{enclave} environment, and data source websites. Therefore, it achieves isolation between network communication and the execution of \textsc{TC}'s core program. Even if the \texttt{Relay} module is attacked, or the network communication packets are tampered, it will not change the normal function of \textsc{TC}. \textsc{TC} system provides a robust security model for the smart contracts' off-chain data interaction, and it has already been launched online as a public service~\cite{30}.

\section{Future Directions}
\label{sec:future}

Based on the above systematic examination on the security of current blockchain systems, we list a few future directions to stir up research efforts into this area. First, nowadays the most popular consensus mechanism used in blockchain is PoW. However, a major disadvantage of PoW is the waste of computing resources. To solve this problem, Ethereum is trying to develop a hybrid consensus mechanism of PoW and PoS. Conducting researches and developing more efficient consensus mechanisms will make a significant contribution to the development of blockchain. Second, with the growth of the number of feature-rich dAPPs, the privacy leakage risk of blockchain will be more serious. A dAPP itself, as well as the process of communication between the dAPP and Internet, are both faced with privacy leakage risks. There are some interesting techniques that can be applied in this problem: code obfuscation, application hardening, execution trusted computing (e.g., Intel SGX), etc. Third, the blockchain will produce a lot of data, including block information, transaction data, contract bytecodes, etc. However, not all of the data stored in blockchain is valid. The smart contract can erase its code by executing \texttt{SUICIDE}/\texttt{SELFDESTRUCT}. In addition, there are a lot of smart contracts containing totally the same code in Ethereum, and many smart contracts are never be executed after their deployments. An efficient data cleanup and detection mechanism is desired to improve the execution efficiency of blockchain systems.

\section{Conclusion}
\label{sec:conclusion}

In this paper, we focus on the security issues of blockchain technology. By studying the popular blockchain systems (e.g., Ethereum, Bitcoin, Monero, etc.), we conduct a systematic examination on the security risks to blockchain. For each risk or vulnerability, we analyze its causes and possible consequence. Furthermore, we survey the real attacks on the blockchain systems, and analyze the vulnerabilities exploited in these attacks. Finally, we summarize blockchain security enhancements and suggest a few future directions in this area.

\section*{Acknowledgements}
This work is supported in part by the National Science Foundation of China (No.61471228, No.61402080, No.61572115, No.61502086, No.61572109), the Key Project of Guangdong Province Science \& Technology Plan (No.2015B020233018), and China Postdoctoral Science Foundation founded project (No.2014M562307).

\bibliographystyle{elsarticle-num}

\bibliography{RefBib}

\end{document}